\documentclass[pra,reprint, showpacs, a4paper]{revtex4-1}
\usepackage{graphicx}
\usepackage{amsmath}
\usepackage{siunitx}
\usepackage{amssymb}
\usepackage{color}
\usepackage[latin1]{inputenc}
\usepackage[normalem]{ulem}
\newcommand{\bichro}[2]{($#1\omega$:$#2\omega$)}
\begin{document}

\title{Atomic photoionization dynamics in ultrashort cycloidal laser fields}

\author{T. Bayer, Ch. Philipp, K. Eickhoff and M. Wollenhaupt}
\affiliation{Carl von Ossietzky Universit\"at Oldenburg, Institut f\"ur Physik, Carl-von-Ossietzky-Stra\ss e 9-11, D-26129 Oldenburg Germany}

\date{\today}

\begin{abstract}
We present numerical simulations of ultrafast multiphoton ionization dynamics in a two-dimensional atomic model driven by co- and counterrotating circularly polarized single-color and bichromatic carrier envelope phase (CEP) stable ultrashort laser pulse sequences. Taking into account phase variations due to CEP fluctuations and the Gouy phase, our results accurately reproduce recently measured photoelectron momentum distributions \cite{Pengel:2017:PRL:053003, Pengel:2017:PRA:043426,Kerbstadt:2019:NC:658,Kerbstadt:2019:APX:1672583}. The time evolution of the complex-valued electron wave function in coordinate and momentum space is calculated to study the bound state- and the vortex formation dynamics. The non-vanishing azimuthal probability current density proves the vortex nature of 
%the $c_7$ rotationally symmetric electron wave packet 
electron wave packets with odd-numbered rotational symmetry. Their angular momentum expectation value assumes half-integer values of $3.5$ (corotating) and $0.5$ (counterrotating). Knowledge of the wave function allows us to analyze the photoionization dynamics and to validate the physical pictures proposed in previous experimental studies. As an outlook, we investigate how electron vortices develop from the multiphoton- to the tunnel regime with increasing laser intensity. 
\end{abstract}

\maketitle

\section{Introduction} \label{sec:intro}
Recently, vortex states of light and matter characterized by an intrinsic orbital angular momentum, such as twisted light beams \cite{Hernandez-Garcia,Shen:2019:LSA:1} and electron vortices \cite{Lloyd:2017:RMP:035004,Bliokh:2017:PR:1}, have attracted much attention. The use of counterrotating circularly polarized (CRCP) time-delayed attosecond pulses was proposed as an alternative way to generate vortex-shaped photoelectron wave packets in the photoionization of helium atoms \cite{NgokoDjiokap:2015:PRL:113004}. Subsequent theoretical studies have refined our understanding of electron vortices by investigating photoionization with intense \cite{Li:2018:OE:878,Kong:2018:JOSAB:2163,Li:2019:IEEEPJ:1} and bichromatic \cite{NgokoDjiokap:2016:PRA:013408,Yuan:2016:PRA:053425,Yuan:2017:JPB:124004, Kong:2018:JOSAB:2163} fields, double ionization \cite{NgokoDjiokap:2017:PRA:013405,Djiokap:2017:JO:124003,Djiokap:2018:PRA:063407} and molecular photoionization \cite{Yuan:2016:PRA:053425,Yuan:2017:JPB:124004,Djiokap:2018:PRA:063407,Chen:2020:PRA:033406}. Control of electron vortices by the carrier envelope phase (CEP) \cite{Yuan:2016:PRA:053425,Li:2019:CPL:063201}, the relative optical phase \cite{Djiokap:2017:JO:124003,Zhen:2020:CPL:136885} and the polarization state \cite{NgokoDjiokap:2016:PRA:013408,Yuan:2017:JPB:124004,NgokoDjiokap:2017:PRA:013405,Li:2017:COL:120202,Li:2018:IEEEPJ:1} was discussed. In addition, vortices have been used as a test bed to validate powerful emerging theoretical techniques, such as the $R$-matrix theory with time dependence \cite{Clarke:2018:PRA:053442,Armstrong:2019:PRA:063416,Brown:2019:CPC:107062}. The first experimental realization of vortex-shaped electron wave packets by single-color multiphoton ionization (MPI) of potassium atoms with CRCP femtosecond laser pulses was reported in \cite{Pengel:2017:PRL:053003,Pengel:2017:PRA:043426}. Subsequently, the generation and manipulation of photoelectron vortices with odd-numbered rotational symmetry by MPI of sodium atoms with bichromatic CRCP and corotating circularly polarized (COCP) cycloidal laser fields was demonstrated in \cite{Kerbstadt:2019:NC:658,Kerbstadt:2019:APX:1672583}. \\
In this contribution we present a numerical study of atomic MPI dynamics with cycloidal femtosecond laser pulses based on the solution of the time-dependent Schr\"odinger equation (TDSE) in two dimensions. To this end, we employ established numerical tools \cite{Tannor:2007:662,Bauer:2017}, which are especially suited for the simulation of photoelectron momentum distributions (PMDs) from atomic MPI with cycloidal femtosecond laser pulses. Since photoionization with CRCP or COCP pulses cancels the cylindrical symmetry of the electron wave function, 2D numerical simulations are necessary to describe the asymmetric polarization profiles but also sufficient to capture the essential features of the PMD. Two-dimensional approaches to solve the TDSE (2dTDSE) have successfully been applied, amongst others, to model the scattering of free particles \cite{Galbraith:1984:AJP:60}, atomic double ionization \cite{Lein:2000:PRL:4707,Doerr:2000:OE:111}, coupled electron-nuclear dynamics in molecules \cite{Shin:1995:JCP:9285,Erdmann:2004:JCP:9666a,Falge:2011:JCP:184307} and the photoelectron circular dichroism \cite{Paul:2018:PRL:233202b}. Very recently, Zhen \textit{et al.} \cite{Zhen:2020:CPL:136885} applied the 2dTDSE method to study electron vortices in the photoionization of hydrogen atoms by CRCP pulse sequences. In that work, the authors examined the influence of the time delay, the optical phases and the wavelength on the properties of the final PMD and discussed their findings in the context of previous theoretical predictions. In contrast, we investigate the vortex formation dynamics by exploring the time evolution of the wave function and analyze recent experimental results on photoelectron vortices \cite{Pengel:2017:PRL:053003,Pengel:2017:PRA:043426,Kerbstadt:2019:NC:658,Kerbstadt:2019:APX:1672583}. The complete picture of the time evolution of the wave function in coordinate and momentum space, provided by our 2dTDSE simulation, reveals the initial field-driven dynamics and the subsequent field-free time evolution of the electron wave packet up to the PMD, which is measured in the experiment. The analysis of the transient bound state populations assesses the perturbative character of the scenarios and reveals the role of intermediate resonances opening up new experimental perspectives for the utilization of photoelectron vortices. The complex-valued wave function provides the phase information from which the probability current density, the topological charge and the expectation value of the azimuthal angular momentum are derived and interpreted. In the region of overlapping free electron wave packets from three- and four-photon ionization the latter assumes half-integer values of $l_z=0.5$ and $l_z=3.5$ for MPI with CRCP and COCP pulse sequences, respectively. To compare our simulation results with previous measurements, we consider the phase variations due to CEP fluctuations and the Gouy phase. \\ 
The 2dTDSE approach and the perturbative MPI (pMPI) description reported in the experimental contributions \cite{Pengel:2017:PRL:053003,Pengel:2017:PRA:043426,Kerbstadt:2019:NC:658,Kerbstadt:2019:APX:1672583} provide complementary information on the MPI dynamics. In the pMPI model, the relevant MPI pathways are determined by selection rules and resonances known from the excitation and ionization scheme. However, multi-photon dipole moments are generally not readily accessible. The relevant physical mechanism of coherent control, i.e. in this case the multipath interference, is already incorporated into the model. This feature of the pMPI model facilitates the predictive design of new coherent control scenarios. In contrast, the 2dTDSE model provides the time evolution of the wave function allowing to extract all observables and derived quantities such as the probability current and the topological charge. It gives access to arbitrary intensity regimes, i.e. perturbative multiphoton ionization, non-perturbative MPI, resonance-enhanced MPI (REMPI), above-threshold ionization (ATI) and tunneling ionization. By combining the complementary information of both approaches, we are able to validate the assumptions of the pMPI description and use our \textit{ab initio} calculation to extend the analysis of experimental data beyond the validity of the pMPI model. \\
The paper is structured as follows: first we present our 2dTDSE model for photoionization with cycloidal femtosecond laser pulses along with the numerical methods in Sec.~\ref{sec:wavepacket propagation}. Then we discuss the simulation results for perturbative and non-perturbative MPI with single color CRCP pulse sequences (Sec.~\ref{sec:results:singlecolor}) and MPI with bichromatic COCP and CRCP pulse sequences (Sec.~\ref{sec:results:bichromatic}) and compare the calculated PMDs to recently reported experimental results. In Sec.~\ref{sec:results:perspectives}, we study how electron vortices develop from the multiphoton- to the tunnel regime as the laser intensity increases.

\section{Wave packet propagation}\label{sec:wavepacket propagation}

\subsection{Theoretical description}\label{theory}

\begin{figure}[htb]
	\includegraphics[width=\linewidth]{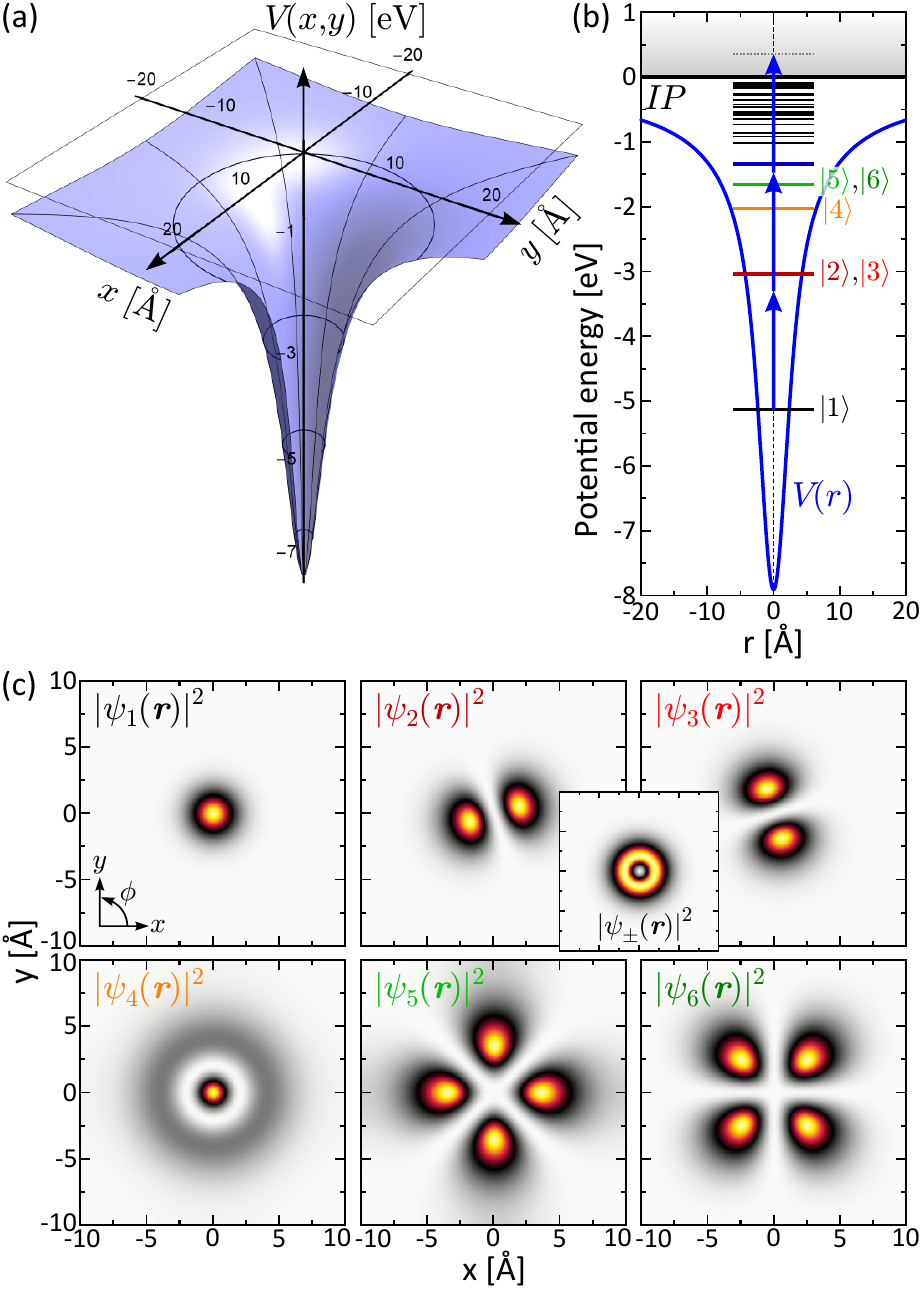}
	\caption{(Color online.) Two-dimensional model to simulate atomic MPI with cycloidal fields.
	(a) Potential $V(x,y)$ of the Sodium-type single active electron atom with screened Coulomb interaction. (b) Scheme for three-photon ionization with $\lambda= \SI{675}{nm}$ laser pulses along with the potential $V(r)$ and the eigenenergy spectrum. The six lowest energy eigenstates $\vert n \rangle $ are numbered. The corresponding densities $|\psi_n(\boldsymbol{r})|^2$ of the eigenstates and the superposition states $|\psi_{\pm}(\boldsymbol{r})|^2$ (inset) are plotted in (c).}
	\label{fig:fig1}
\end{figure}

The experiments which we analyze were performed on different atomic systems (potassium and sodium). The underlying physical mechanisms, however, are universal. Therefore, and for consistency, all numerical simulations presented below were performed on the same 2D atomic model system. The model mimics a sodium-type single active electron atom with screened Coulomb interaction described by the potential \cite{Sprik:1988:JCP:1592,Shin:1995:JCP:9285,Erdmann:2004:JCP:9666a}
\begin{equation}\label{eq:potential}
	V(r)=-\frac{z e^2}{4\pi\varepsilon_0}\frac{\mathrm{erf}(r/a)}{r}.
\end{equation}
This type of potential was originally introduced by Sprik \textit{et al.} \cite{Sprik:1988:JCP:1592} and extensively used by H. Metiu \cite{Shin:1995:JCP:9285,Shin:1996:JPC:7867}, V. Engel \cite{Erdmann:2004:JCP:9666a,Schaupp:2019:JCP:164110} and S. Gr\"afe \cite{Falge:2011:JCP:184307,Falge:2017:PCCP:19683} to study coupled electron-nuclear dynamics in molecular systems. By choice of $z=0.9048$ for the effective nuclear charge and $a=\SI{1.8544}{\AA}$ for the screening parameter, the model reproduces the ionization potential (IP) $\varepsilon_{ip}=\SI{5.139}{eV}$ of sodium and the prominent $3p\rightarrow3s$ sodium D-line at $\lambda_D=\SI{589}{nm}$, i.e. $\varepsilon_{D}=\SI{2.104}{eV}$ \cite{Kramida:2018} (SI units are used throughout this paper). The potential is shown in Fig.~\ref{fig:fig1}(a). Due to its hyperbolic long-range behavior, it exhibits a manifold of high-lying Rydberg states (see Fig.~\ref{fig:fig1}(b)) which are important for the realistic modeling of multiphoton excitation. In total, we obtain 37 bound electronic states. In addition, the potential features a continuum of free electronic states suitable to describe photoionization of the atom, in contrast to the solution $V(r)\propto\ln(r/r_0)$ of the 2D Poisson equation, which diverges asymptotically. \\
We consider the interaction of the atom with a polarization-shaped femtosecond laser field consisting of two components with individual central frequencies $\omega_i$ ($i=1,2$), amplitudes $\mathcal{E}_i$, time delays $\tau_i$, relative phases $\varphi_i$ and polarization states, but with a common CEP $\varphi_{ce}$ and the same Gaussian-shaped pulse envelope $g(t)=e^{-\ln(4)(\frac{t-\tau_i}{\Delta t})^2}$ of a duration defined by the intensity full width at half maximum $\Delta t$:
\begin{align}\label{eq:e_field}
	\boldsymbol{E}^+(t) = \;& \mathcal{E}_1\,g(t-\tau_1)\,e^{i(\omega_1 t+\varphi_1+\varphi_{ce})}\,\boldsymbol{e}_{q_1} \notag \\
	& \quad+\;\mathcal{E}_2\,g(t-\tau_2)\,e^{i(\omega_2 t+\varphi_2+\varphi_{ce})}\,\boldsymbol{e}_{q_2}.
\end{align}
Herein, $\boldsymbol{e}_{q_i}$ ($q_i=-1,+1$) are the spherical unit vectors $\boldsymbol{e}_{-1}=-(\boldsymbol{e}_x-i\boldsymbol{e}_y)/\sqrt{2}$ and $\boldsymbol{e}_1=-(\boldsymbol{e}_x+i\boldsymbol{e}_y)/\sqrt{2}$ describing right- (RCP) and left-handed (LCP) circular polarization, respectively. Laser fields of the form Eq.~\eqref{eq:e_field} have been discussed in detail in the supplementary information of \cite{Kerbstadt:2019:NC:658}. The real-valued laser electric field is obtained as $\boldsymbol{E}(t)=\Re[\boldsymbol{E}^+(t)]$. \\
To calculate the PMD, which is the central observable in the experiment, we start by propagating the TDSE
\begin{equation}\label{eq:tdse}
	i\hbar\frac{\partial}{\partial t}\psi(\boldsymbol{r},t)=\mathcal{H}(\boldsymbol{r},t)\psi(\boldsymbol{r},t).
\end{equation}
on a discrete 2D spatial grid (for computational details see Sec.~\ref{sec:model}). The single active electron Hamiltonian in the dipole approximation is given by $\mathcal{H}(\boldsymbol{r},t)=-\frac{\hbar^2}{2m_e}\Delta+V(r)+e\,\boldsymbol{r}\cdot\boldsymbol{E}(t)$, with $m_e$ and $-e$ being the electron mass and charge, respectively. The propagated wave function on a discrete equidistant temporal grid $\{t_j\}_{j=1,..,J}$ with step size $\delta t=t_{j+1}-t_j$ reads \cite{Cohen-Tannoudji:1977:3}
\begin{equation}\label{eq:psi_prop}
	\psi(\boldsymbol{r},t_{j+1})=e^{-i\frac{\delta t}{\hbar} \mathcal{H}(\boldsymbol{r},t_j)}\,\psi(\boldsymbol{r},t_j).
\end{equation}
Initially, we assume the atom to be in the ground state, i.e., $\psi(\boldsymbol{r},t_1)=\psi_1(\boldsymbol{r})$ (cf. Sec.~\ref{sec:model}). The time evolution operator on the right-hand side of Eq.~\eqref{eq:psi_prop} is evaluated using a Fourier-based split operator method \cite{Feit:1982:JCP:412a} 
\begin{equation}\label{eq:t_evo_op}
	e^{-i\frac{\delta t}{\hbar} \mathcal{H}(\boldsymbol{r},t_j) } \approx e^{-i\frac{\delta t}{2\hbar} \mathcal{V}(\boldsymbol{r},t_j)} \, \mathcal{F}^{-1} \, e^{- i \frac{\hbar\delta t k^2}{2m_e} } \, \mathcal{F} \, e^{-i\frac{\delta t}{2\hbar} \mathcal{V}(\boldsymbol{r},t_j)},
\end{equation}
where $\mathcal{V}(\boldsymbol{r},t)=V(r)+e\,\boldsymbol{r}\cdot\boldsymbol{E}(t)$ is the effective time-dependent potential, $\mathcal{F}$ denotes the 2D Fourier transformation with respect to $\boldsymbol{r}$ and $\hbar k=\hbar|\boldsymbol{k}|$ is the modulus of the electron momentum. Subsequent to the interaction, i.e., after the decay of the laser pulse, the wave function is propagated field-free with the same time step until its free part has propagated sufficiently far from the bound part of the wave function to apply a radially symmetric sigmoidal splitting filter $f(r)$ similar to \cite{Heather:1987:JCP:5009}. The filter separates the wave function into an outer part $\psi_f(\boldsymbol{r},t_J)=f(r)\,\psi(\boldsymbol{r},t_J)$, describing the outgoing photoelectron wave packet, and an inner part $\psi_b(\boldsymbol{r},t_J)=[1-f(r)]\,\psi(\boldsymbol{r},t_J)$ still bound by the interaction potential. To minimize reflections at the boundaries of the numerical grid, we use absorbing boundary conditions \cite{Kosloff:1986:JCP:363,Santra:2006:PRA:034701} implemented by an imaginary polynomial potential $U(r)=-iU_0\,(r/r_0)^{2n}$ of the order $2n=16$, with $U_0=\SI{0.69}{eV}$ and $r_0=x_{max}$ (cf. Sec.~\ref{sec:model}). Other groups have used transparent boundary conditions for this purpose \cite{Antoine:2008:CCP:729,Feshchenko:2013:PRE:053308}. In the experiment, the density of the free part is measured by a detector situated far away from the interaction region \cite{Wollenhaupt:2009:APB:647,Pengel:2017:PRL:053003,Kerbstadt:2019:NC:658}. On the way to the detector, the coordinate space wave function $\psi_f(\boldsymbol{r},t)$ spreads, due to the quadratic dispersion relation $\omega(\boldsymbol{k})=\hbar \boldsymbol{k}^2/(2m_e)$ of free matter waves, and evolves asymptotically into its own Fourier transform \cite{Wollenhaupt:2002:PRL:173001,Winter:2006:OC:285,Wollenhaupt:2013:CPC:1341,Goda:2013:NP:102} (see also Appendix~\ref{app:asymptotics}):
\begin{equation}\label{eq:rho_asymptotic}
	\lim_{t\rightarrow\infty} |\psi_f(\boldsymbol{r},t)|^2\propto |\tilde{\psi}_f(\boldsymbol{k},t_0)|^2,
\end{equation}
where $t_0$ denotes the beginning of the field-free time evolution. Therefore, the experiment yields the PMD of the free electron wave packet even though the detector measures electrons in coordinate space. Numerically, we determine the PMD as
\begin{equation}
	\mathcal{P}(\boldsymbol{k})\propto\left|\mathcal{F}[\psi_f(\boldsymbol{r},t_0)](\boldsymbol{k})\right|^2,
\end{equation}
at a time step $t_0$ after the decay of the laser pulse, when the free part of the wave function has propagated sufficiently far to be separated from the bound part, but not too far to reach the absorbing boundaries. The simulation then allows us to compare the calculated PMD to the measured PMD and, in particular, to 'watch' the PMD arise asymptotically in the shape of the coordinate space probability density $\varrho(\boldsymbol{r},t)=|\psi_f(\boldsymbol{r},t)|^2$. Moreover, with the wave function at hand, we are able to extract quantities related to its phase such as the probability current density (cf. Eq.~\eqref{eq:current}) and the topological charge.
\begin{figure*}[t]
	\includegraphics[width=.9\linewidth]{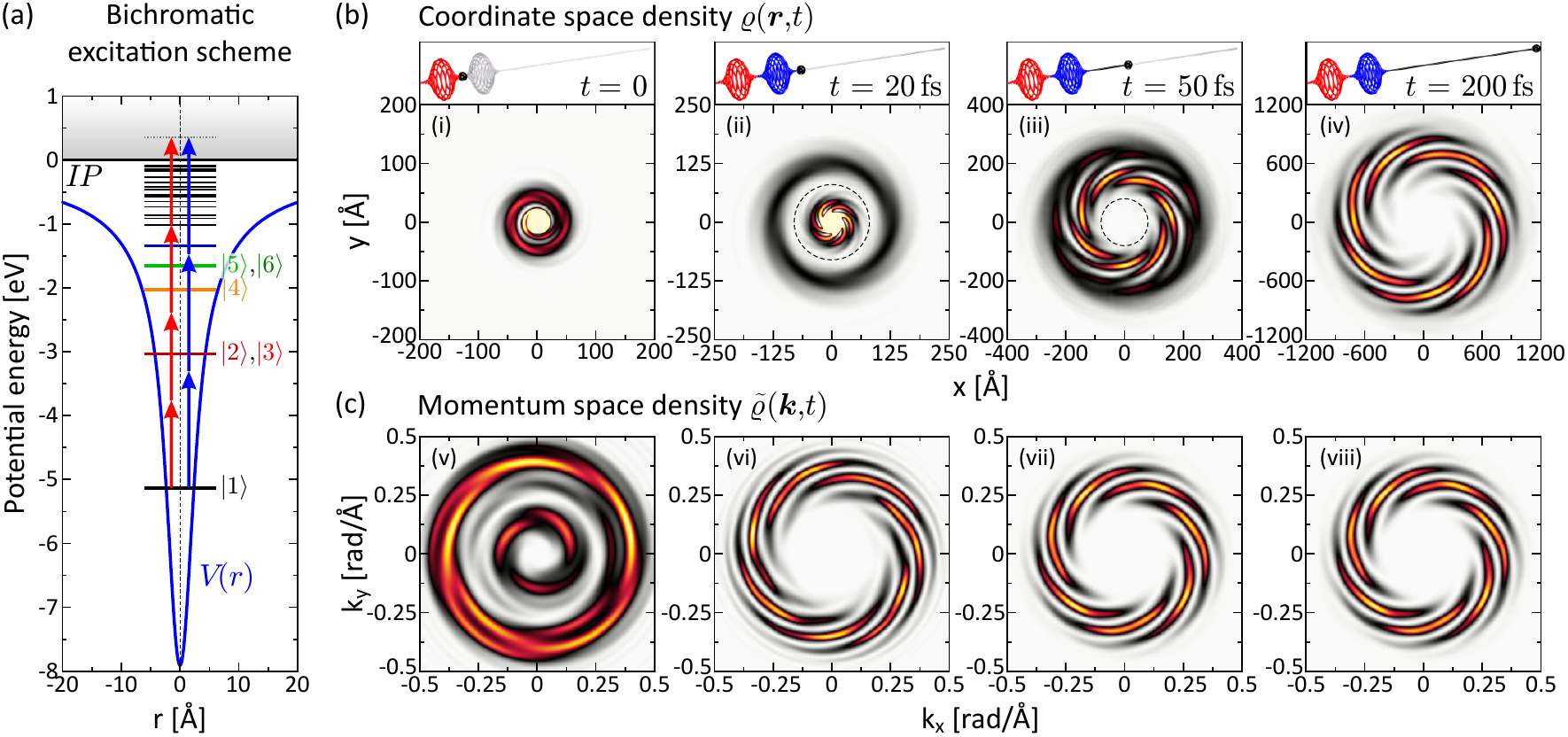}
	\caption{(Color online.) Time evolution of $c_7$ vortex from perturbative three- vs. four-photon ionization. (a) shows the corresponding bichromatic excitation scheme. The resulting density of free electron wave packet is shown in coordinate space (b) and momentum representation (c). The pulse parameters of the Gaussian-shaped bichromatic \bichro{3}{4} CRCP double pulse sequence as defined in the text are: $\Delta t=\SI{7.5}{fs}$, $\lambda_1=\SI{900}{nm}$, $\lambda_2=\SI{675}{nm}$ $\tau_1=-\SI{10}{fs}$, $\tau_2=+\SI{10}{fs}$, $\mathcal{E}_1=\SI{7.1e6}{V/cm}$ (\SI{8.3e11}{W/cm^2}), $\mathcal{E}_2=\SI{3.2e6}{V/cm}$ (\SI{2.6e11}{W/cm^2}). The top insets sketches the time evolution of the laser electric field.}
	\label{fig:fig2}
\end{figure*}

\subsection{Numerical Model}\label{sec:model}

The TDSE is solved numerically on a discrete square spatial grid $[-x_{max},x_{max}-\delta x]\times[-y_{max},y_{max}-\delta y]$ with $x_{max}=y_{max}=\SI{500}{\AA}$. Each dimension is sampled by $2^{10}$ points. The corresponding step size is $\delta x=\delta y\approx\SI{1}{\AA}$. The time propagation of the wave function is performed in two stages. The laser-atom interaction is treated in the first stage. To this end, we sample the laser field on a symmetric temporal grid $[-t_{max},t_{max}-\delta t]$ by $2^{13}$ points, using a step size of $\delta t\approx\SI{10}{as}$. The wave function $\psi(\boldsymbol{r},t_j)$ is then calculated according to Eq.~\eqref{eq:psi_prop}. Subsequently, the wave function is propagated field-free on the time grid $[t_{max},t_J]$ with the same step size $\delta t$ and on the same spatial grid. For investigations of the long-term evolution of $\psi(\boldsymbol{r},t)$ in Figs.~\ref{fig:fig2} and \ref{fig:fig3}, the field-free time evolution was performed on enhanced spatial grids ranging to a maximum of $x_{max}=y_{max}=\SI{3500}{\AA}$ and sampled by up to $3300$ points per dimension.\\
Knowledge of the resonances of the atomic system is vital for both the design and the analysis of the different MPI scenarios discussed in Sec.~\ref{sec:results}. The bound state eigenenergies $\varepsilon_n$ and corresponding eigenfunctions $\psi_n(\boldsymbol{r})$ are determined by diagonalization of the unperturbed Hamiltonian
\begin{equation}
	\mathcal{H}_0=-\frac{\hbar^2}{2m_e}\Delta+V(r)=-\frac{\hbar^2}{2m_e}\mathcal{F}^{-1}(-ik)^2\mathcal{F}+V(r)
\end{equation}
using the Fourier grid Hamiltonian method \cite{Marston:1989:JCP:3571}. The eigenspectrum of the bound system is shown in Fig.~\ref{fig:fig1}(b). Figure~\ref{fig:fig1}(c) illustrates the densities $|\psi_n(\boldsymbol{r})|^2$ of the first six eigenstates and the superposition states $|\psi_{\pm}(\boldsymbol{r})|^2 = |\psi_2(\boldsymbol{r})\pm i \psi_3(\boldsymbol{r})|^2/2$. For computational reasons, the eigenfunctions were initially determined on a smaller grid ($x_{max}=y_{max}=\SI{20}{\AA}$) than the grid used for the TDSE propagation. To ensure a highly accurate initial state, we extended the ground state wave function $\psi_1(\boldsymbol{r})$ to the full grid by zero padding, and refined the result by imaginary-time propagation \cite{Tal-Ezer:1986:CPL:223}. Convergence is typically achieved in less than 100 iterations. To analyze the laser-induced transient dynamics of the neutral system, we calculate the time-dependent bound state populations
\begin{equation}
	p_n(t)=|\langle\psi_n|\psi_b(t)\rangle|^2=\left|\iint\limits_{-\infty}^{\quad+\infty} \psi_n^\ast(\boldsymbol{r})\psi_b(\boldsymbol{r},t)\,d^2\boldsymbol{r}\,\right|^2.
\end{equation}

\section{Results}\label{sec:results}

\begin{figure}[tb]
	\includegraphics[width=\linewidth]{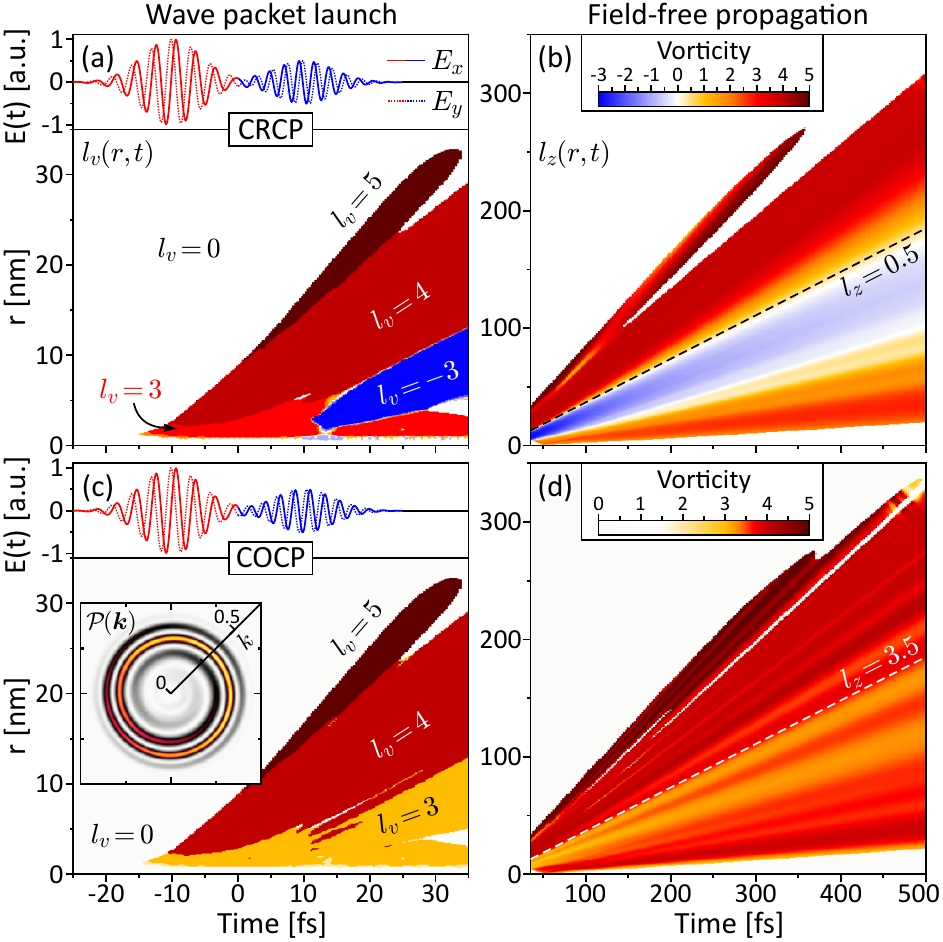}
	\caption{(Color online.) Transient $r$-resolved topological charge of [(a) and (b)] the $c_7$ vortex from bichromatic MPI with CRCP pulses (cf. Fig.~\ref{fig:fig2}) and [(c) and (d)] a $c_1$ vortex from bichromatic MPI with COCP pulses under otherwise identical conditions. Note the different color scale in (c) and (d). The PMD of the latter is shown in the inset to (c). The dashed lines in the long-term evolution (b) and (d) indicate the center-of-mass motion of the corresponding free electron wave packets. The topological charge of the $c_7$ vortex converges asymptotically towards $l_z=0.5$. The asymptotic value in the case of the $c_1$ vortex is $l_z=3.5$.}
	\label{fig:fig3}
\end{figure}

In this section, we present our simulation results and compare them to recent experimental observations published in \cite{Pengel:2017:PRL:053003,Pengel:2017:PRA:043426,Kerbstadt:2019:NC:658,Kerbstadt:2019:APX:1672583}. Besides the reproduction of the PMD, the simulation model allows us to analyze the transient dynamics of additional physical quantities such as the bound electron dynamics (Sec.~\ref{sec:results:singlecolor}), the spatiotemporal evolution of the released electron wave packets (Fig.~\ref{fig:fig2}), the transient topological charge (Fig.~\ref{fig:fig3}) and the relative orientation between final PMD and laser field (Sec.~\ref{sec:results:bichromatic}). Those quantities are generally not observed in experiments but are invaluable for the characterization of the quantum state, thus providing deeper insights into the physical mechanisms at play. In particular, we are able to validate the physical picture employed in the experimental papers \cite{Pengel:2017:PRL:053003,Pengel:2017:PRA:043426,Kerbstadt:2019:NC:658,Kerbstadt:2019:APX:1672583}. After the analysis of the experimental results, we apply the 2dTDSE model to study new scenarios by venturing into different intensity regimes (Sec.~\ref{sec:results:perspectives}). \\
We start by examining the time evolution of a free electron wave packet from its birth to its asymptotic state, to illustrate the relation between the coordinate and momentum space representation of its wave function (see Appendix~\ref{app:asymptotics} for a mathematical description). As an example, we consider the perturbative bichromatic three- vs. four-photon ionization scenario discussed in Sec.~\ref{sec:results:bichromatic}, which yields a seven-armed free electron vortex created by a CRCP \bichro{3}{4} pulse sequence. The bichromatic excitation scheme is depicted in Fig.~\ref{fig:fig2}(a). The pulse parameters are provided in the figure caption. The numerical results are presented in (b) and (c). The top row shows the coordinate space density $\varrho(\boldsymbol{r},t)$ (note the different grid ranges of frames $(i)$-$(iv)$), the lower row shows the corresponding momentum space density $\tilde{\varrho}(\boldsymbol{k},t)=|\tilde{\psi}_f(\boldsymbol{k},t)|^2$. The pulse shapes are indicated in the top insets of frames $(i)$-$(iv)$ in (b). Snapshots of the dynamics are displayed for four different instants of the time evolution. The first snapshot is taken at $t=0$, after the decay of the first (LCP, red) pulse and before the onset of the second (RCP, blue) pulse. The saturated spot at the origin of frame $(i)$ is the bound part of the density (cf. Fig.~\ref{fig:fig1}(c)). Only a tiny fraction of the density is released and spirals outwards as a first partial wave. The momentum space density in $(v)$ is obtained by application of a narrow spatial splitting filter $f(r)$. However, because bound and free part of the wave function still overlap significantly, the result has to be taken \textit{cum grano salis}. The second snapshot shown in $(ii)$ is taken at $t=\SI{20}{fs}$, after the decay of the blue pulse. In the meantime, the first partial wave has departed, moving radially at the mean group velocity $v_g=\frac{\hbar k_0}{m_e}$, where $\hbar k_0$ is the central component of the PMD in radial direction. The second partial wave, launched by the blue pulse, arises in the shape of a six-armed vortex. This structure is due to the interference with a bound wave packet created by the red pulse via resonant three-photon excitation of the low-lying Rydberg states (see Fig.~\ref{fig:fig2}(a)). The corresponding momentum space density $(vi)$, however, has already developed a pronounced \emph{seven}-armed vortex shape. The vortex has a clockwise sense of rotation when following its arms outwards. At the instant of the third snapshot, at $t=\SI{30}{fs}$, the fast components of the second partial wave catch up with the slow components of the first \cite{Wollenhaupt:2005:ARPC:25,Winter:2006:OC:285}. Both wave packets commence to overlap. Their interference now produces a seven-armed spiral structure in the coordinate space density $(iii)$. Meanwhile, both partial waves have traveled sufficiently far from the bound part to apply the splitting filter -- indicated by dashed circles in $(ii)$ and $(iii)$ -- without disturbance from the bound states. The resulting momentum space density $(vii)$ reveals already the final, stationary PMD. Upon further (field-free) propagation, the momentum space wave function acquires only the dispersion phase $\omega(\boldsymbol{k})t$ (cf. Eq.~\eqref{eq:psi_free}) leaving its momentum distribution $\tilde{\varrho}(\boldsymbol{k},t)$ unaltered. As the overlap of both partial waves increases, the same shape emerges in the coordinate space density, as discussed in Appendix~\ref{app:asymptotics}. Already at $t=\SI{200}{fs}$, the instant of the fourth snapshot, the coordinate space density $(iv)$ is almost indistinguishable from the PMDs $(vii)$ and $(viii)$. Asymptotically, $\varrho(\boldsymbol{r},t)$ continues to expand and disperse but its overall shape is maintained. 
 \\
The availability of the complex-valued wave function $\psi_f(\boldsymbol{r},t)$ provides additional opportunities to characterize and visualize the vortex formation. For example, its phase $\chi(\boldsymbol{r},t)=\arg[\psi_f(\boldsymbol{r},t)]$ allows us to calculate and examine the probability current density \cite{Cohen-Tannoudji:1977:3}
\begin{align}\label{eq:current}
\boldsymbol{j}(\boldsymbol{r},t) & = \frac{\hbar}{2m_ei}\left[\psi_f(\boldsymbol{r},t)\boldsymbol{\nabla}\psi_f^\ast(\boldsymbol{r},t)-\psi_f^\ast(\boldsymbol{r},t)\boldsymbol{\nabla}\psi_f(\boldsymbol{r},t)\right]\notag \\
& = \frac{\hbar}{m_e}\Im\left[\psi_f^\ast(\boldsymbol{r},t)\boldsymbol{\nabla}\psi_f(\boldsymbol{r},t)\right] \notag \\
& = \frac{\hbar}{m_e}\varrho(\boldsymbol{r},t)\boldsymbol{\nabla}\chi(\boldsymbol{r},t).
\end{align}
Vortex structures in the current field of a wave function have long been recognized as 'vortex states' in quantum physics \cite{Bialynicki-Birula:2000:PRA:032110,Macek:2009:PRL:143201,Ovchinnikov:2010:PRL:203005,Velez:2018:PRA:043421,Lloyd:2017:RMP:035004,Bliokh:2017:PR:1}. Such states arise around phase singularities, i.e., points where the phase of a wave function is indeterminate -- accompanied by zero amplitude -- and feature a non-vanishing azimuthal component of the current density. In contrast to the free electron vortices discussed in \cite{NgokoDjiokap:2015:PRL:113004,Pengel:2017:PRL:053003}, which exhibit a spiral-shaped probability \emph{density}, vortex states are determined by a helical topology of the \emph{phase}. This topology is characterized by an integer winding number
\begin{equation}\label{eq:topo_charge}
l_v = \frac{1}{2\pi}\ointop_{\mathcal{C}}\boldsymbol{\nabla}\arg[\psi(\boldsymbol{r})]\cdot d\boldsymbol{r}, 
\end{equation}
termed vorticity or topological charge, describing the phase $2\pi l_v$ accumulated along a contour $\mathcal{C}$ enclosing the singularity. To study the formation of the $c_7$ vortex in Fig.~\ref{fig:fig2}(b), we introduce the transient, radially resolved topological charge $l_v(r,t)$ by evaluating Eq.~\eqref{eq:topo_charge} for every time step $t$ and integrating along concentric circles $\mathcal{C}=\mathcal{C}(r)$ with radius $r$. Thus, Eq.~\eqref{eq:topo_charge} becomes
\begin{equation}\label{eq:topo_charge_diff}
l_v(r,t) = \frac{1}{2\pi}\intop_0^{2\pi}\frac{\partial}{\partial\phi}\chi(r,\phi,t)d\phi,
\end{equation}
where $\chi(r,\phi,t)$ is the transient phase of the wave function $\psi_f(r,\phi,t)$ in polar representation. The result is presented in Fig.~\ref{fig:fig3}. The initial time evolution of the $c_7$ vortex is shown in (a). Extended white areas with $l_v=0$ result from phase blanking in regions of negligible density. The dark-red area describes the first partial wave launched around $t=\SI{-10}{fs}$ by the LCP red pulse. The left and right edges evolve linearly with different slopes, indicating the constant outward motion and broadening of the wave packet. Its topological charge of $\l_v=4$ is consistent with the absorption of four LCP photons and, hence, ionization into a continuum with angular momentum $|g,m_\ell=4\rangle$. Simultaneously, the red area with $l_v=3$ is created but remains localized at small $r$, i.e., close to the ionic core. This part of the wave function is the $|f,3\rangle$-type Rydberg wave packet resonantly excited by the red pulse via absorption of three LCP photons. The steep finger with $l_z=5$ is attributed to a weak ATI contribution created by the first pulse. The blue area describes the second partial wave launched by the time-delayed blue pulse around $t=\SI{+10}{fs}$. Its topological charge is $l_v=-3$, corresponding to the ionization into an $|f,-3\rangle$-type continuum by absorption of three RCP photons. The topological charges of the two partial waves in the $l_v$-map are strictly separated by a distinct line along which both waves have equal amplitudes. The discrete plateau-like separation of different $l_v$-areas is inherent to the vorticity definition in Eq.~\eqref{eq:topo_charge} and persists even as the overlap between both waves grows until they are no longer distinguishable. A continuous and hence more physical definition of the vorticity is given by the expectation value of the orbital angular momentum $\ell_z$ (in units of $\hbar$):
\begin{equation}\label{eq:topo_charge_lz}
l_z=\langle \ell_z/\hbar\,\rangle=\frac{1}{i}\iint\limits_{-\infty}^{\quad+\infty} \psi^\ast(\boldsymbol{r})\frac{\partial}{\partial\phi}\psi(\boldsymbol{r})d^2\boldsymbol{r}.
\end{equation}
Instead of the volume integration, we integrate only over the angular coordinate to obtain a radius- and time-resolved expression analogous to Eq.~\eqref{eq:topo_charge_diff} 
\begin{equation}\label{eq:topo_charge_diff_lz}
l_z(r,t)=\frac{1}{i\mathcal{N}(r,t)}\intop_0^{2\pi} \psi_f^\ast(r,\phi,t)\frac{\partial}{\partial\phi}\psi_f(r,\phi,t)d\phi,
\end{equation}
with the normalization factor $\mathcal{N}(r,t)=\int_0^{2\pi}|\psi_f(r,\phi,t)|^2d\phi$. This definition is better suited to investigate the long-term evolution of the vorticity as it smoothly accounts for the continuous merging of partial waves with different topological charges. The numerical results are presented in Fig.~\ref{fig:fig3}(b). For large times, the distinct separation of the dark-red and the blue $l_z$-area is gradually lifted due to the growing overlap of the two partial waves. The black dashed line indicates the center-of-mass motion of the total wave packet. Along this line, the topological charge converges to a half-integer value of $l_z\approx0.5$. The same value is obtained by performing the full volume integration in Eq.~\eqref{eq:topo_charge_lz} for every time $t>\SI{35}{fs}$, implying that the overall topological charge of the free electron wave packet $\psi_f(\boldsymbol{r},t)$ rapidly converges towards its final value. This final value agrees with the result $l_{z,\mathrm{pMPI}}=(4-3)/2=0.5$ predicted by the pMPI model \cite{Kerbstadt:2019:APX:1672583}. \\
Figures~\ref{fig:fig3}(c) and (d) show the numerical result obtained for COCP pulses, i.e., when both the red and the blue pulse are LCP. In this case, the blue pulse launches a partial wave with a topological charge of $l_v=3$ as indicated by the yellow area in (c). As a result, the total wave packet $\varrho(\boldsymbol{r},t)$ evolves asymptotically into a one-armed free electron vortex displayed in the inset to (c). In this limit, the topological charge tends towards a half-integer value of $l_z\approx3.5$, as indicated by the white dashed line in (d). This result coincides again with the value of $l_{z,\mathrm{pMPI}}=(4+3)/2=3.5$ predicted by the pMPI model.

\subsection{Single Color MPI}\label{sec:results:singlecolor}

\begin{figure}[t]
	\includegraphics[width=\linewidth]{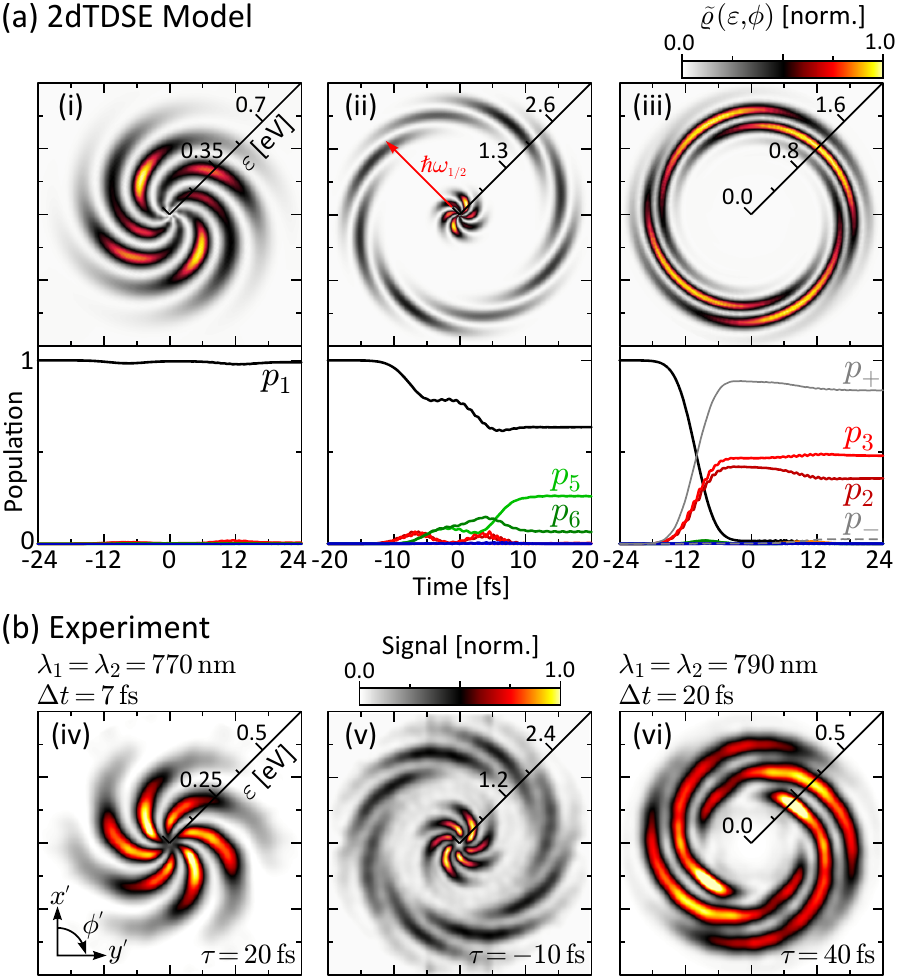}
	\caption{(Color online.) Free electron vortices from three-photon ionization with single color CRCP pulse sequences. Numerical simulation results in (a) are compared to experimental results in (b) (taken from \cite{Pengel:2017:PRA:043426}). The simulations are based on sequences of CRCP $\Delta t=\SI{7.5}{fs}$ pulses with $(i)$ $\lambda_1=\lambda_2=\SI{675}{nm}$, $\tau_{1/2}=\mp\SI{10}{fs}$, $\mathcal{E}_{1/2}=\SI{3.5e6}{V/cm}$, $(ii)$ $\lambda_1=\lambda_2=\SI{675}{nm}$, $\tau_{1/2}=\pm\SI{5}{fs}$, $\mathcal{E}_{1/2}=\SI{1.8e7}{V/cm}$ and $(iii)$: $\lambda_1=\lambda_2=\SI{589}{nm}$, $\tau_{1/2}=\mp\SI{10}{fs}$, $\mathcal{E}_{1/2}=\SI{1.5e7}{V/cm}$. The bottom frames in (a) show the corresponding bound state population dynamics indicating $(i)$ perturbative interaction, $(ii)$ semi-perturbative interaction and $(iii)$ resonant $\pi$-pulse excitation.}
	\label{fig:fig4}
\end{figure}

Next, we investigate three different MPI scenarios for the creation of free electron vortices with even-numbered rotational symmetry. The scenarios are designed along the lines of experiments recently reported in \cite{Pengel:2017:PRL:053003,Pengel:2017:PRA:043426}, based on single color MPI using CRCP pulse sequences. The numerical results are presented in Fig.~\ref{fig:fig4}(a) and compared to the corresponding measurements in (b). We start with a perturbative three-photon ionization scenario initially demonstrated by Pengel \textit{et al.} on potassium atoms \cite{Pengel:2017:PRL:053003}. The excitation scheme adapted to our sodium-type model is depicted in Fig.~\ref{fig:fig1}(b). The pulse sequence consists of an LCP first pulse followed by an RCP second pulse with a time delay of $\tau=\tau_2-\tau_1=\SI{20}{fs}$. Both pulses have a duration of $\Delta t=\SI{7.5}{fs}$ and a field amplitude of $\mathcal{E}_{1/2}=\SI{3.5e6}{V/cm}$ corresponding to a peak intensity of $I_{1/2}=\SI{4.6e11}{W/cm^2}$. Similar to the experiment, the central wavelength $\lambda_{1/2}=\SI{675}{nm}$ was chosen such that the field ionizes the atom by absorption of three photons, producing photoelectrons close to the ionization threshold. Because the single color MPI is insensitive to the CEP \cite{Pengel:2017:PRL:053003,Pengel:2017:PRA:043426} all optical phases were set to zero, i.e., $\varphi_{ce}=\varphi_1=\varphi_2=0$. The time evolution of the released free electron wave packet is analogous to the one presented in Fig.~\ref{fig:fig2}. Two partial waves are launched successively by the two time-delayed pulses. During the subsequent free propagation they disperse, merge and interfere. In this case, however, the total density $\varrho(\boldsymbol{r},t)$ evolves asymptotically into a six-arm electron vortex, i.e., a vortex-shaped wave packet with (approximate) $c_6$ rotational symmetry. The corresponding PMD in energy representation is displayed in frame $(i)$, showing excellent agreement with the measurement presented in frame $(iv)$. The calculated $c_6$ vortex is energetically centered at $\varepsilon=\SI{0.34}{eV}$ in good agreement with the expected photoelectron excess energy of $\varepsilon=3\hbar\omega_{1/2}-\varepsilon_{ip}=\SI{0.37}{eV}$. Examination of the bound state population dynamics, depicted in the bottom panel of $(i)$, reveals that the ground state population $p_1(t)\equiv 1$ remains essentially unchanged throughout the interaction, confirming the perturbative character of the scenario. \\
In the next step, we study the free electron vortex created by ATI as initially demonstrated in \cite{Pengel:2017:PRA:043426}. For this purpose, we increased the amplitude of both pulses by a factor of 5 to $\mathcal{E}_{1/2}=\SI{1.8e7}{V/cm}$ (\SI{1.1e13}{W/cm^2}). In addition, in order to match the experimental settings \cite{Pengel:2017:PRA:043426}, we decreased the time delay to $\tau=\SI{-10}{fs}$ and changed its sign so that the RCP pulse precedes the LCP pulse. The simulation result is shown in frame $(ii)$ and compared to the measurement in $(v)$. Besides the $c_6$ vortex near the ionization threshold, both yield an ATI contribution centered at about one photon energy $\hbar\omega_{1/2}$ (red arrow) above the threshold contribution. The latter are likewise vortex-shaped but exhibit a $c_8$ rotational symmetry as discussed in \cite{Pengel:2017:PRA:043426}. Both the threshold and the ATI vortices rotate counterclockwise due to the reversed pulse ordering. The simulated ATI contribution was enhanced by a factor of 500 to increase its visibility. Similarly, the measured ATI signal was enhanced by a factor of 4 \cite{Pengel:2017:PRA:043426}. Qualitatively, the simulation agrees well with the experiment. Deviations in the exact shape of the contributions are essentially due to differences in the related optical spectra. The ultrabroad white light supercontinuum used in the experiment is highly structured with a bandwidth supporting the generation of pulses as short as $\Delta t=\SI{5}{fs}$. The spectrum used in the simulation is Gaussian-shaped and chosen narrower to avoid the intermediate $|1\rangle\rightarrow|2\rangle$ resonance (cf. Fig.~\ref{fig:fig1}(b)). In the simulation, the ATI peak is centered at $\varepsilon=\SI{2.06}{eV}$. Compared to the expected position of \SI{2.21}{eV}, the signal is shifted downwards by \SI{150}{meV} due to the dynamic Stark effect. As revealed by the bound state population dynamics, the pulse sequence transfers some population from the ground state to the $d$-type bound sates $|5\rangle$ and $|6\rangle$. These states are nearly 2-photon resonant with the laser spectrum (cf. Fig.~\ref{fig:fig1}(c)), however, with a relatively large red-detuning of \SI{182}{meV}. The population in these states indicates a significant AC-Stark shift at play in the bound system, rationalizing the observed red-shift of the ATI signal. \\
The third single color MPI scenario discussed here is based on resonant non-perturbative excitation of the atom using a sequence of $\pi$-pulses (with respect to the $|1\rangle\rightarrow|2\rangle$ resonance). It was first shown in \cite{Pengel:2017:PRL:053003,Pengel:2017:PRA:043426} that 1+2 REMPI of potassium atoms using $\pi$-pulses yields a free electron vortex with $c_4$ rotational symmetry. A typical measurement result is depicted in frame $(vi)$. To reproduce the measurement by our simulation, we tuned the central wavelength to the sodium $|1\rangle\rightarrow|2\rangle$ resonance at $\lambda_{1/2}=\SI{589}{nm}$. In addition, we changed the time delay back to $\tau=\SI{20}{fs}$. Using a field amplitude of $\mathcal{E}_{1/2}=\SI{1.5e7}{V/cm}$ (\SI{7.7e12}{W/cm^2}), we obtained the high-contrast four-armed electron vortex shown in frame $(iii)$. Analogous to the experiment, the simulated $c_4$ vortex rotates clockwise due to the pulse ordering. The simulated vortex is centered at $\varepsilon=\SI{1.19}{eV}$ which agrees well with the expected position of \SI{1.18}{eV}. The measured $c_4$ vortex is centered at a smaller kinetic energy of $\varepsilon=\SI{0.35}{eV}$ due to the different resonance energy in the potassium atom relative to the IP. Regarding the relevant features, however, the simulation agrees well with the measurement. Analysis of the population dynamics reveals, that the pulse area of both pulses is indeed $\pi$. The LCP first pulse transfers almost the entire population from the ground state to a superposition of the resonant states $|2\rangle$ and $|3\rangle$. More precisely, the population is transferred efficiently to the spherical state $|+\rangle=(|2\rangle+i|3\rangle)/\sqrt{2}$, depicted in Fig.~\ref{fig:fig1}(c), in accordance with the dipole selection rule $\Delta m=+1$ for LCP excitation. At the same time, the orthogonal state $|-\rangle=(|2\rangle-i|3\rangle)/\sqrt{2}$ remains unpopulated. The corresponding populations $p_\pm(t)$ are depicted as solid and dashed gray lines, respectively, in the bottom panel of $(iii)$. Complete inversion of the resonant subsystem $\{|1\rangle,|2/3\rangle\}$ is not accomplished, because some population escapes to the high-lying states $|9\rangle$ ($s$-type), $|12\rangle$ and $|13\rangle$ (both $d$-type) in two-photon resonance. \\
The agreement between the PMDs obtained numerically by \textit{ab initio} propagation of the TDSE and the experimental results validates the intuitive pMPI model presented in \cite{Pengel:2017:PRL:053003,Pengel:2017:PRA:043426}. This model treats the excitation and the ionization of the atom separately assuming perturbative conditions for the latter \cite{Wollenhaupt:2003:PRA:015401}. Especially the accurate reproduction of the non-perturbative $c_4$ vortex in the 2dTDSE simulation by forward design of the light field confirms the model interpretation.

\begin{figure}[htb!]
	\includegraphics[width=1 \linewidth]{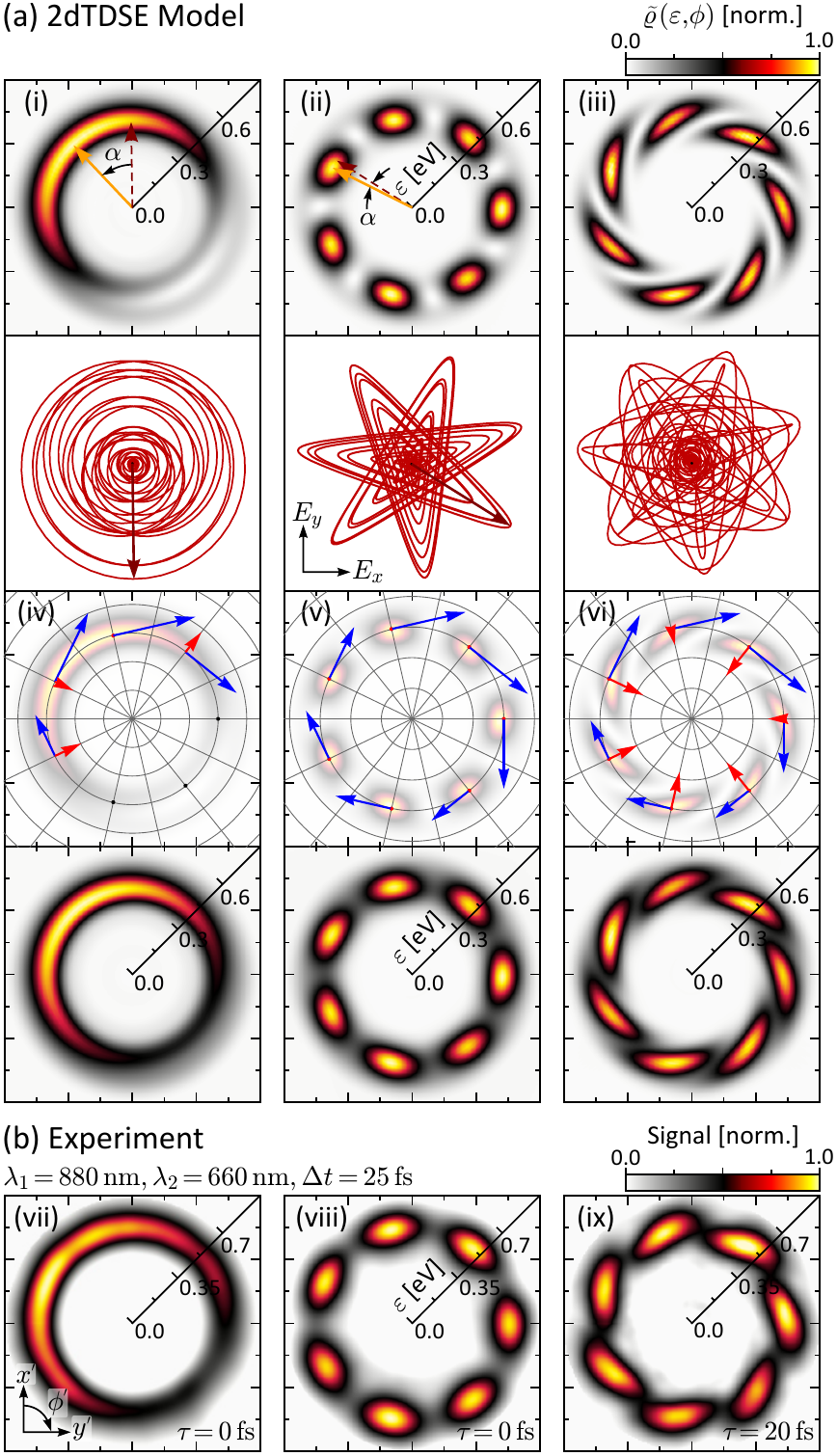}
	\caption{(Color online.) Free electron wave packets from three- vs. four-photon ionization with bichromatic \bichro{3}{4} pulse sequences. Numerical results in (a) are compared to experimental results in (b) (taken from \cite{Kerbstadt:2019:NC:658}). The simulation pulse parameters are $\Delta t=\SI{30}{fs}$, $\lambda_1=\SI{900}{nm}$, $\lambda_2=\SI{675}{nm}$, $\mathcal{E}_1=\SI{4.8e6}{V/cm}$ and $\mathcal{E}_2=\SI{2.7e6}{V/cm}$. Scenarios $(i)$ and $(ii)$ are based in time-overlapping COCP and CRCP sequences, respectively, with $\tau=0$. Scenario $(iii)$ builds on a time-delayed CRCP sequence with $\tau=\SI{20}{fs}$. The second row displays the corresponding pulse polarization profiles. The third row shows the calculated probability current density $\boldsymbol{j}(\varepsilon,\phi)$ decomposed into its radial (red) and azimuthal (blue) component. In the rotationally averaged PMD $\langle \mathcal{P}(\boldsymbol{k})\rangle_{\varphi_{opt}}$, shown in the bottom row of (a), phase variations due to CEP fluctuations and the Gouy phase are taken into account.}
	\label{fig:fig5}
\end{figure}

\subsection{Bichromatic MPI}\label{sec:results:bichromatic}

In this section, we study the creation of free electron wave packets with odd-numbered rotational symmetry. It was shown theoretically \cite{NgokoDjiokap:2016:PRA:013408,Douguet:2016:PRA:033402,Yuan:2016:PRA:053425,Yuan:2017:JPB:124004,Douguet:2016:PRA:033402} and demonstrated experimentally \cite{Eckart:2016:PRL:133202,Mancuso:2016:PRA:053406,Kerbstadt:2019:NC:658} that such wave packets are created by \emph{bichromatic} MPI using commensurable frequency CRCP or COCP fields. Three different MPI scenarios, which were recently reported in \cite{Kerbstadt:2019:NC:658}, are discussed. The scenarios are based on the interference of three- vs. four-photon ionization pathways in sodium atoms excited by \bichro{3}{4} pulse sequences with different polarization states. The corresponding excitation scheme is depicted in Fig.~\ref{fig:fig2}(a). Similar to the experiment, the central wavelengths were adjusted to $\lambda_1=\SI{900}{nm}$ and $\lambda_2=3\lambda_1/4=\SI{675}{nm}$ and the pulse duration was set to $\Delta t=\SI{30}{fs}$. The pulses were chosen relatively long, i.e., spectrally narrow to avoid spectral overlap of the relevant photoelectron contribution -- from three- vs. four-photon ionization -- with energetically adjacent contributions from intrapulse frequency mixing \cite{Kerbstadt:2019:APX:1672583}. The field amplitude of the blue pulse was set to $\mathcal{E}_2=\SI{2.7e6}{V/cm}$ to ensure perturbative interaction conditions (cf. Sec.~\ref{sec:results:singlecolor}). The field amplitude of the red pulse was adapted to $\mathcal{E}_1=\SI{4.8e6}{V/cm}$ by matching the total photoelectron yield produced by each single color alone, thus maximizing the three- vs. four-photon interference contrast of the two-color signal. The corresponding peak intensities are $I_1=\SI{8.3e11}{W/cm^2}$ and $I_2=\SI{2.6e11}{W/cm^2}$. Since the rotation of the PMD by the optical phases was already investigated in detail in the experiment \cite{Kerbstadt:2019:NC:658}, they were initially set to zero in the simulations. The numerical results are presented in Fig.~\ref{fig:fig5}(a) and compared to corresponding measurements in (b). The first bichromatic MPI scenario is based on a temporally overlapping ($\tau=0$) COCP \bichro{3}{4} field consisting of two LCP colors. The calculated PMD shown in frame $(i)$ displays a single spherical standing wave pattern, i.e., with one cycle per full angle, reminiscent of a crescent. The simulation result agrees well with the measurement shown in $(vii)$. The lobe of the measured PMD is slightly broadened in azimuthal ($\phi$) direction due to residual CEP averaging in the experiment, which we will address below. The corresponding laser field $\boldsymbol{E}(t)$ is plotted parametrically in the second row of (a). In the COCP case, it exhibits a cycloidal, almost heart-shaped polarization profile. The dark-red arrow indicates the maximum field amplitude $\boldsymbol{E}_0=\boldsymbol{E}(0)$. According to the pMPI model, the maximum of the photoemission is expected in the opposite direction (dashed arrow in $(i)$). However, the simulation reveals a counterclockwise rotation of the crescent (yellow arrow) with respect to $-\boldsymbol{E}_0$ by about $\alpha=43^\circ$. Changing the circularity of the pulse inverts the sign of this rotation (not shown). In \cite{Eickhoff:2020:PRA:013430} it was shown that the relative orientation between the crescent and the field is additionally determined by quantum phases $\varphi_{qm}$ acquired during the ionization process. In this case, the ionization phase arises due to the three-photon Rydberg resonance in the ionization pathway of the red pulse. To verify this assumption, we performed two separate studies in which we varied (1) the central wavelengths of both colors, scanning the third order spectrum of the red pulse across the resonance, and (2) the spectral bandwidth of both colors for slightly off-resonant excitation. In both cases, we found a strong sensitivity of the rotation angle $\alpha$ to the spectral pulse parameters. We also verified that the interaction is perturbative so that additional strong-field-induced phase contributions due to the dynamic Stark effect are negligible. \\
The second bichromatic MPI scenario is based on a time-overlapping CRCP \bichro{3}{4} pulse, generated by changing the circularity of the blue component from LCP to RCP. The corresponding laser field, shown below frame $(ii)$, exhibits a seven-armed star-shaped polarization profile. The resulting PMD is a spherical standing wave with seven cycles per full angle, i.e., with $c_7$ rotational symmetry. The simulation result is in good accordance with the experiment shown in $(viii)$. Again, the seven lobes in the measurement are slightly broader due to phase averaging (see below). In the simulation, we observe a small counterclockwise rotation of the PMD relative to $-\boldsymbol{E}_0$ (dark-red arrow). As in the COCP case, the rotation direction is inverted by exchanging the circularities of the two colors (not shown). The origin of the rotation is the same quantum phase $\varphi_{qm}$ imparted by the Rydberg resonance to the partial wave packet created by the red pulse. The rotation angle $\alpha=6^\circ$, however, is only about $1/7$ of the rotation in the COCP case. This is readily explained by the pMPI model, from which the rotation angle was derived as \cite{Kerbstadt:2019:NC:658}
\begin{equation}\label{eq:rotation}
\alpha=\frac{\varphi_{opt}+\varphi_{qm}}{4\mp 3}.
\end{equation}
The negative (positive) sign in the denominator applies to the COCP (CRCP) case. The optical phase $\varphi_{opt}=\varphi_{ce}+4\varphi_2-3\varphi_1$ is zero here. Equation~\eqref{eq:rotation} hence yields a ratio of $1/7$ between the quantum-phase-induced PMD rotations in the COCP and CRCP case. In general, the ionization phase rotates the standing wave pattern by a specific fraction of a cycle, irrespective of the angular frequency (symmetry). Therefore, the lower the rotational symmetry of the PMD the more sensitive is the absolute rotation to additional quantum phases. \\
In the third bichromatic MPI scenario, we introduced a time delay of $\tau=\SI{20}{fs}$ between the two colors of the CRCP \bichro{3}{4} field. As a result, the rotational symmetry of the field is lifted. However, the symmetry of the created PMD shown in $(iii)$ is preserved. Due to the time delay, the the lobes of the $c_7$ standing wave tilt energetically, yielding a clockwise rotating free electron vortex with seven arms (see also Fig.~\ref{fig:fig2}), in good accordance with the measurement in $(ix)$. A remarkable feature of the vortex with odd-numbered symmetry is its probability current density $\boldsymbol{j}(\varepsilon,\phi)$ plotted in frame $(vi)$. All current densities were calculated according to Eq.~\eqref{eq:current} but using the spectral wave function $\tilde{\psi}_f(\varepsilon,\phi)$ instead of the coordinate space wave function. This replacement is justified, since asymptotically both wave functions differ only by a radial phase (cf. Eq.~\eqref{eq:psi_asymptotic}). Radial phases resulting from the general outwards propagation of the wave packets were eliminated to highlight the subtle intrinsic contributions imprinted by the ionization process, i.e., the laser field. It has been shown for the pMPI model that for three- vs. four-photon ionization by \bichro{3}{4} laser pulse sequences with a time delay $\tau$, the current density of the resulting photoelectron wave packet (in the polarization plane) is given by \cite{Kerbstadt:2019:APX:1672583} 
\begin{equation}\label{eq:current_bichro}
\boldsymbol{j}(\varepsilon,\phi) = \frac{\hbar}{2m_e} \tilde{\varrho}(\varepsilon,\phi) \left(-\frac{\tau}{\hbar} \boldsymbol{e}_{\varepsilon} \mp \frac{4\pm 3}{\varepsilon} \boldsymbol{e}_{\phi} \right). 
\end{equation}
The positive (negative) sign in the numerator of the azimuthal component applies to the COCP (CRCP) case, while the overall sign of this component distinguishes between an LCP (minus) and RCP (plus) red pulse. According to Eq.~\eqref{eq:current_bichro}, the radial component vanishes for $\tau=0$. This is confirmed by the current densities in frames $(iv)$ and $(v)$ obtained for temporally overlapping colors. In these cases, we observe an almost strictly azimuthal current density (blue arrows) indicating a probability flux in $-\phi$-direction. The clockwise rotation of the flux is in agreement with $-(4+3)=-7$ for the COCP and $-(4-3)=-1$ for the CRCP case. For better comparison, the current density in the COCP case was scaled by a factor of $1/7$. The resulting similarity of the azimuthal flux in the COCP and the CRCP case implies that also the magnitude ratio is described accurately by Eq.~\eqref{eq:current_bichro}. In the time-delayed vortex scenario, however, the calculated current density exhibits a pronounced radial component in addition (red arrows). As predicted by Eq.~\eqref{eq:current_bichro}, the radial component points inwards indicating a flux towards smaller kinetic energies. We conclude that the non-vanishing azimuthal and radial parts of the current density prove the vortex nature of the $c_7$ rotationally symmetric PMD both in the hydrodynamic formulation of quantum mechanics \cite{Madelung:1927:ZP:322,Bialynicki-Birula:2000:PRA:032110,Macek:2009:PRL:143201} and in the intuitive picture introduced by Djiokap \textit{et al.} \cite{NgokoDjiokap:2015:PRL:113004}. Hence, the probability current density is a suitable quantity to characterize both notions of quantum vortices. \\
Finally, we discuss the effect of variations in the optical phases on the observed PMD. We consider two sources of phase variations, i.e. (1) CEP fluctuations of the laser radiation of about $\sigma_{ce}=\SI{200}{mrad}$ \cite{Kerbstadt:2019:NC:658} and (2) the variation of the Gouy phase of about $\sigma_{Gouy}=\SI{800}{mrad}$ along the laser intensity profile $I^4(z)$ on the propagation axis \cite{Born:1999:1,Hoff:2017:NP:947}. Equation~\eqref{eq:rotation} describes the rotation of the PMD due to the optical phase $\varphi_{opt}$ by an angle of $\alpha = \varphi_{opt}/n$ with $n=1$ in the COCP case and $n=7$ in the CRCP case. Therefore, we take the effect of the phase variation into account in our simulation by rotational averaging over the PMD using a Gaussian distribution function $p_{\sigma}(\varphi_{opt})$ with an RMS of $\sigma^2=\sigma^2_{ce}+\sigma^2_{Gouy}$
\begin{equation}\label{eq:rotationalaveraging}
\langle \mathcal{P}(\boldsymbol{k})\rangle_{\varphi_{opt}} = \intop_{-\infty}^{\infty} p_{\sigma}(\varphi_{opt}) \, \mathcal{P}(\boldsymbol{k},\varphi_{opt}/n) \, d\varphi_{opt}.
\end{equation}
The averaged PMDs shown in the fourth row of Fig.~\ref{fig:fig5} are almost indistinguishable from the experimental results. In our case the major part of the phase variation is introduced by the Gouy phase rather than the CEP fluctuations. Because the Gouy phase is determined by the Rayleigh range, further reduction of the phase variations can only be achieved in higher-order non-linear processes confining the $n^{th}$-order intensity profile $I^n(z)$. The comparison of simulated and measured data emphasizes the importance of CEP stabilization in bichromatic experiments: in the presence of the Gouy phase, measurements with insufficiently stabilized CEP ($>\SI{1}{rad}$) would wash out the PMD, resulting in a torus-shaped angular distribution similar to an incoherent superposition of the PMDs from both colors.

\subsection{Perspectives}\label{sec:results:perspectives}

\begin{figure}[t]
	\includegraphics[width=\linewidth]{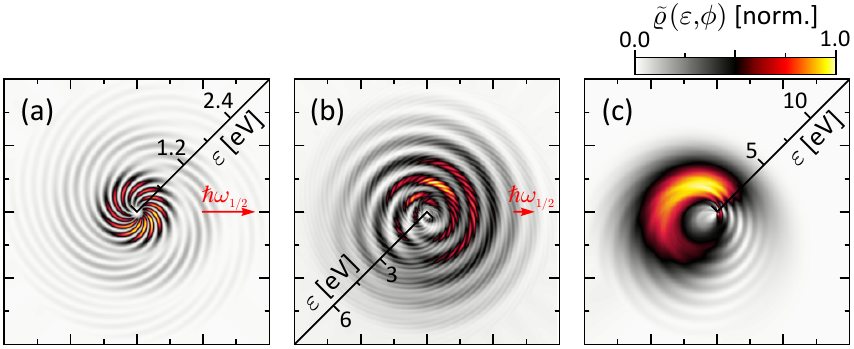}
	\caption{(Color online.) Free electron vortices from strong-field ionization of the sodium-model by intense mid-infrared single color CRCP sequences with a central wavelength of $\lambda_{1/2}=\SI{1300}{nm}$, single pulse duration of $\Delta t=\SI{7.5}{fs}$ and a time delay of $\tau=\SI{20}{fs}$. The peak amplitudes are (a) $\mathcal{E}_{1/2}=\SI{3.5e7}{V/cm}$ (\SI{4.6e13}{W/cm^2}), (b) $\SI{1.1e8}{V/cm}$ (\SI{4.1e14}{W/cm^2}) and (c) $\SI{1.8e8}{V/cm}$ (\SI{1.1e15}{W/cm^2}). The PMDs capture the transition from the ATI to the tunneling ionization regime.}
	\label{fig:fig6}
\end{figure}

The excellent agreement of our numerical results with previous experimental findings validates the 2dTDSE approach and suggests its application to unexplored regimes. The simulation can be used to predict characteristic experimental signatures and thus serve as a tool for the design of new experiments such as the vortex formation dynamics in the presence of electron-electron interactions \cite{NgokoDjiokap:2017:PRA:013405,Djiokap:2017:JO:124003} or coupled electron-nuclear dynamics \cite{Yuan:2016:PRA:053425,Yuan:2017:JPB:124004,Djiokap:2018:PRA:063407,Chen:2020:PRA:033406}. So far, the physical properties and ionization dynamics of free electron vortices created by tunneling ionization are still unexplored. Although vortices from non-perturbative MPI have been investigated both experimentally \cite{Pengel:2017:PRL:053003,Pengel:2017:PRA:043426} and theoretically \cite{Li:2018:OE:878,Kong:2018:JOSAB:2163,Li:2019:IEEEPJ:1} the transition of photoelectron vortices from the ATI to the tunneling regime has not been examined. In Fig.~\ref{fig:fig6}(a)-(c), we present a series of free electron vortices from strong-field ionization of the sodium-type model for increasing field amplitude. To promote the tunneling mechanism, we used intense \emph{mid-infrared} ($\lambda_{1/2}=\SI{1300}{nm}$, i.e. $\SI{0.95}{eV}$) few-cycle ($\Delta t=\SI{7.5}{fs}$) CRCP pulse sequences ($\tau=\SI{20}{fs}$) in these simulations. The relative phase between the two pulses was set to $\Delta\varphi=\varphi_2-\varphi_1=\pi$. The PMD shown in (a) was obtained using a field amplitude of $\mathcal{E}_{1/2}=\SI{3.5e7}{V/cm}$ which corresponds to a peak intensity of $I_{1/2}=\SI{4.6e13}{W/cm^2}$. It displays an extended clockwise-rotating vortex pattern composed of a 12-arm vortex at the ionization threshold (center) -- resulting from 6-photon ionization -- seamlessly adjoined by a weaker 14-arm vortex from ATI, which itself merges into the barely discernible 16-arm vortex from second order ATI. The latter was slightly enhanced by a factor of 1.5 for better visibility. This field amplitude marks the onset of the ATI regime. By increasing the amplitude to $\mathcal{E}_{1/2}=\SI{1.1e8}{V/cm}$ (\SI{4.1e14}{W/cm^2}), we obtained the result shown in (b) -- note the different energy scale. In this intermediate intensity regime, the PMD is characterized by multiple overlapping ATI orders forming an intricate vortex structure. The filigree clockwise rotating vortex pattern inherent to each individual ATI ring is superimposed by an overall counterclockwise rotating vortex shape resulting from the interference of adjacent ATI orders. Upon further increase of the field amplitude to $\mathcal{E}_{1/2}=\SI{1.8e8}{V/cm}$ (\SI{1.1e15}{W/cm^2}), we reach the onset of the tunneling regime (Keldysh parameter $\gamma=0.6$). The resulting PMD is shown in (c). At this field strength, the first pulse already ionizes the atom completely. The released photoelectron wave packet is driven ponderomotively in the continuum by the second pulse, similar to photoelectron (angular) streaking \cite{Itatani:2002:PRL:173903,Kienberger:2004:Nature:817,Eckle:2008:NP:565} known from attosecond physics. Hence the deflection of the PMD from the center, which increases its already pronounced asymmetry. We found that the streaking deflection is highly sensitive to the relative phase $\Delta\varphi$. In this intensity regime, we did not obtain regular vortex patterns. Due to the bound state depletion by the first pulse, the vortex formation mechanism is disabled and no free electron vortex is observed in the tunneling regime under the conditions considered in this paper. A closer analysis and, in particular, a further advance into the tunneling regime will be subject of a forthcoming contribution.

\section{Conclusion}\label{sec:conclusion}

Motivated by recent experiments on photoelectron vortices \cite{Pengel:2017:PRL:053003,Pengel:2017:PRA:043426,Kerbstadt:2019:NC:658,Kerbstadt:2019:APX:1672583}, we have studied atomic multiphoton ionization (MPI) with corotating (COCP) and counterrotating circularly polarized (CRCP) femtosecond laser pulse sequences by numerically solving the two-dimensional time-dependent Schr\"odinger equation (2dTDSE). We have calculated the time evolution of the electronic wave function in coordinate and momentum space to analyze the transient bound state population dynamics and the subsequent vortex formation dynamics. By considering phase variations due to CEP fluctuations and the Gouy phase, our numerical results accurately reproduced recently measured photoelectron momentum distributions (PMDs). The comparison of simulated and experimental data for ionization with single color CRCP laser pulse sequences \cite{Pengel:2017:PRL:053003} included non-perturbative $c_4$ vortices, perturbative $c_6$ vortices, and $c_8$ electron vortices in the above-threshold ionization (ATI). Simulated electron vortices for bichromatic CRCP and COCP laser pulse sequences, i.e. the odd-numbered $c_7$ vortices and the electron crescent \cite{Kerbstadt:2019:NC:658}, were almost indistinguishable from the experimental results. \\
For our analysis, we used the phase information from the complex-valued wave function $\psi(\boldsymbol{r},t)$ in addition to extract the probability current density $\boldsymbol{j}(\boldsymbol{r},t)$, the time- and radius-dependent topological charge $l_v(r,t)$ and the expectation value of the orbital angular momentum $\langle \ell_z\,\rangle$ which yielded the more physical quantity $l_z(r,t)$. The probability current density $\boldsymbol{j}(\boldsymbol{r},t)$ of the $c_7$ rotationally symmetric electron wave packet showed a non-vanishing azimuthal probability current density confirming its vortex nature, both in the hydrodynamic formulation of quantum mechanics and in the intuitive picture introduced by Djiokap \textit{et al.} \cite{NgokoDjiokap:2015:PRL:113004}. The time evolution of the radius-dependent topological charge was rationalized by identifying regions of a Rydberg wave packet ($l_v=3$), the initial partial free electron wave packet from LCP four-photon ionization ($l_v=4$) and the second partial wave packet from RCP three-photon ionization ($l_v=-3$). Finally, the analysis of the radius-dependent expectation value of the orbital angular momentum reveals a half integer value of $l_z=0.5$ (CRCP) and $l_z=3.5$ (COCP) of the topological charge in the region of overlapping free electron wave packets from three- and four-photon ionization. In this respect, the 2dTDSE approach provides additional information on the MPI dynamics that complements the insights from the perturbative MPI (pMPI) picture. In the pMPI model we make assumptions about the relevant MPI paths motivated by selection rules and resonances and obtain an intuitive understanding of coherent control by multipath interference. In contrast, the 2dTDSE description is model-independent and thus applicable to any ionization regime, including perturbative MPI, non-perturbative MPI, resonance-enhanced MPI, ATI and tunneling ionization. \\
Our model was employed to reproduce and analyze recent experimental measurements. This procedure provided detailed insights into the ionization dynamics. In addition, the excellent agreement with previous experimental findings validated our numerical approach and suggested its application to unexplored regimes. We will use our model as a tool for designing novel experiments by predicting their characteristic signatures. In order to simulate the full three-dimensional momentum distribution measured by photoelectron tomography \cite{Wollenhaupt:2009:APB:647} we will extend the model by an additional spatial dimension. Since photoelectron distributions in general are also affected by spin-orbit coupling \cite{Bayer:2019:NJP:033001}, further refinements will take into account the electron spin. Eventually, the bichromatic coherent control of popular multi-electron systems, such as noble gas atoms \cite{Kerbstadt:2019:EPJWC:07003}, will additionally require the inclusion of electron-electron interactions effects.

\section{acknowledgments}
	
Financial support from the Deutsche Forschungsgemeinschaft via the priority program QUTIF (Program No. SPP1840) is gratefully acknowledged. \\

\appendix

\section{Asymptotic time evolution of the free electron wave function}\label{app:asymptotics}

The discussion below is similar to the treatment presented recently by Winter \textit{et al.} \cite{Winter:2006:OC:285} and Goda \textit{et al.} \cite{Goda:2013:NP:102}, and closely related to the saddle-point approximation method \cite{Reid:1988:SS:213}. After the interaction with the laser pulse, the free electron wave function in coordinate space reads
\begin{equation}
	\psi_f(\boldsymbol{r},t)=\frac{1}{2\pi}\iint \limits_{-\infty}^{\quad+\infty}\tilde{\psi}_f(\boldsymbol{k},t)e^{i\boldsymbol{k\cdot\boldsymbol{r}}}d^2\boldsymbol{k}.
\end{equation}
For ease of notation, we assume the field-free time evolution to start at $t_0=0$. During the propagation, the momentum space wave function $\tilde{\psi}_f(\boldsymbol{k},t)$ of the electron acquires only the dispersion phase $\omega(\boldsymbol{k})t=\frac{\hbar\boldsymbol{k}^2}{2m_e}t$:
\begin{align}\label{eq:psi_free}
	\psi_f(\boldsymbol{r},t) & = \frac{1}{2\pi} \iint \limits_{-\infty}^{\quad+\infty}\tilde{\psi}_f(\boldsymbol{k},0)e^{i\left(\boldsymbol{k\cdot\boldsymbol{r}}-\frac{\hbar t}{2m_e}\boldsymbol{k}^2\right)}d^2\boldsymbol{k} \\
	& = \frac{1}{2\pi}e^{i\frac{m_e}{2\hbar t}\boldsymbol{r}^2}\iint \limits_{-\infty}^{\quad+\infty}\tilde{\psi}_f(\boldsymbol{k},0)e^{-i\frac{\hbar t}{2m_e}\left(\boldsymbol{k}-\frac{m_e}{\hbar t}\boldsymbol{r}\right)^2}d^2\boldsymbol{k} \notag
\end{align}
In the limit $t\rightarrow\infty$, the exponential in the kernel of the integral oscillates rapidly. Only in the vicinity of $\boldsymbol{k}=\frac{m_e}{\hbar t}\boldsymbol{r}$, where its phase vanishes, we expect significant contributions. The momentum space wave function may therefore be replaced by $\tilde{\psi}_f(\frac{m_e}{\hbar t}\boldsymbol{r},0)$. The remaining integral can be solved analytically yielding a factor of $-i\frac{2\pi m_e}{\hbar t}$. Thus, for large times $t$, i.e., for long propagation distances, the free electron wave function can asymptotically be approximated by
\begin{equation}\label{eq:psi_asymptotic}
\psi_f(\boldsymbol{r},t) \approx -\frac{im_e}{\hbar t}\;e^{i\frac{m_e}{2\hbar t}\boldsymbol{r}^2}\;\tilde{\psi}_f\left(\frac{m_e}{\hbar t}\boldsymbol{r},0\right),
\end{equation}
which verifies Eq.~\eqref{eq:rho_asymptotic}. We note, that the above treatment is applicable also to the three-dimensional problem.

\end{document}